\documentclass[preprintnumbers,amsmath,twocolumn,superscriptaddress,amssymb,showpacs]{revtex4}

\usepackage{graphicx}
\usepackage{dcolumn}
\usepackage{bm}

\begin{document}

\title{Extrinsic Curvature, Geometric Optics, and Lamellar Order on Curved Substrates}
\author{Randall D. Kamien}\affiliation{Department of Physics and Astronomy, University of Pennsylvania, Philadelphia PA, 19104}\affiliation{School of Mathematics, Institute for Advanced Study, Princeton, NJ 08540}
\author{David R. Nelson}
\affiliation{Department of Physics, Harvard University, Cambridge MA, 02138}
\author{Christian D. Santangelo}
\affiliation{Department of Physics, University of Massachusetts, Amherst MA, 01003}
\author{Vincenzo Vitelli}
\affiliation{Instituut-Lorentz, Universiteit Leiden, Postbus 9506, 2300 RA Leiden, The Netherlands}

\date{\today}

\begin{abstract}
When thermal energies are weak, two dimensional lamellar structures confined on a curved substrate display complex patterns arising from the competition between layer bending and compression in the presence of geometric constraints. We present broad design principles to engineer the geometry of the underlying substrate so that a desired lamellar pattern can be obtained by self-assembly. Two distinct physical effects are identified as key factors that contribute to the interaction between the shape of the underlying surface and the resulting lamellar morphology. The first is a local ordering field for the direction of each individual layer which tends to minimize its curvature with respect to the three-dimensional embedding. The second is a non-local effect controlled by the intrinsic geometry of the surface that forces the normals to the (nearly incompressible) layers to lie on geodesics, leading to caustic formation as in optics. As a result, different surface morphologies with predominantly positive or negative Gaussian curvature can act as converging or diverging lenses respectively. By combining these ingredients, as one would with different optical elements, complex lamellar morphologies can be obtained. This smectic optometry enables the manipulation of lamellar configurations for the design of novel materials.
\end{abstract}

\maketitle

\section{Introduction}\label{sec:intro}

Though many are taught that there are merely three phases of matter, {\sl solid, liquid,} and {\sl gas}, the understanding of broken symmetries, Nambu-Goldstone modes \cite{Nambu,Goldstone}, and the Anderson-Higgs-Kibble \cite{Anderson,Higgs,Kibble} mechanism allows prediction, control, and elucidation of other novel forms of matter.  Liquid crystalline phases interpolate between the simple forms of matter sketched above.  The nematic has partially broken rotational invariance \cite{Onsager}, the smectic phase has broken one-dimensional translation invariance \cite{deGennessm}, and the hexatic phase \cite{hexatic} has broken two-dimensional rotational invariance.  These materials not only afford a deeper understanding of condensed matter, they (and their lyotropic cousins, diblock copolymers \cite{BatesandFred}), are also of great technological interest, from displays to coatings, from drug delivery to hybrid materials.  Technological applications provide challenges for the theorist and, in particular, pose problems with imperfect boundary conditions and geometries.  In this paper, we focus on striped or smectic phases on \textit{curved} substrates.  Our work is motivated by elegant experiments by Hexemer and Kramer \cite{hexemerthesis} who probed films of diblock copolymers on corrugated surfaces.   Explaining the properties of columnar layers in this geometry is important for controlling the resulting microstructure of the phase on frozen, undulating surfaces.

Lamellar, smectic, striped, and columnar phases can all be treated in a single unified, framework on a two-dimensional surface, with the columns, layers, {\sl etc.} lying {\sl in} the tangent plane.  The physics of smectics is beautiful and intricate:
the coupling between their geometry and rotation invariance ensures the presence of essential nonlinearities in the strain that lead to anomalous elasticity \cite{pelcovits81,pelcovits82} and dramatic departures from linear elasticity even for small strains \cite{brener,santangelo03}.   These nonlinearities make the theory of the nematic-to-smectic-A transition notoriously difficult as well \cite{ntsma} and the effort to capture them has enhanced our understanding of global versus local symmetries.  

Posed with complex boundary conditions, smectics often form focal conic domains \cite{Klemanbook} in which the layers remain equally spaced but acquire a large curvature to accommodate their growth in a confined geometry.  In the past decade or so, smectic phases with cubic order have been discovered \cite{Pansu}, which have pushed our understanding of the competing tendencies of uniform layer spacing and non-vanishing curvature.  Indeed, the two requirements cannot generically be reconciled with the layer topology \cite{didonna02}. This leads to a subtle type of geometric frustration in which there is no local obstruction to finding uniformly-spaced layers but their construction leads to diverging layer curvatures.  These focal conics are, in fact, analogous to the caustic singularities in optics \cite{Nyebook}. 
It has been conjectured that the frustration in smectics can be lifted by introducing curvature in the background space \cite{didonna02}, much as the nematic blue phase is not frustrated on the surface of a four-dimensional sphere \cite{sethna}.  It is, of course, not possible to experimentally study layered systems in high dimensional space.  However,
a simpler but not less subtle form of geometric frustration also exists in two-dimensional smectics lying on curved substrates \cite{hexemerthesis,santangelo07,zhang08}.  In contrast to the case of two-dimensional crystals, \cite{vitelliPNAS,Bow00}
the connection between strain and substrate curvature is indirect, and consequently, configurations with uniformly-spaced layers are numerous. However, the layers nevertheless inherit curvature from the underlying substrate due to compression elasticity. Smectics exhibit quite different phenomena from other ordered phases on curved surfaces and new theoretical ideas are required to understand them.

\begin{figure*}
\includegraphics[width=0.9 \textwidth]{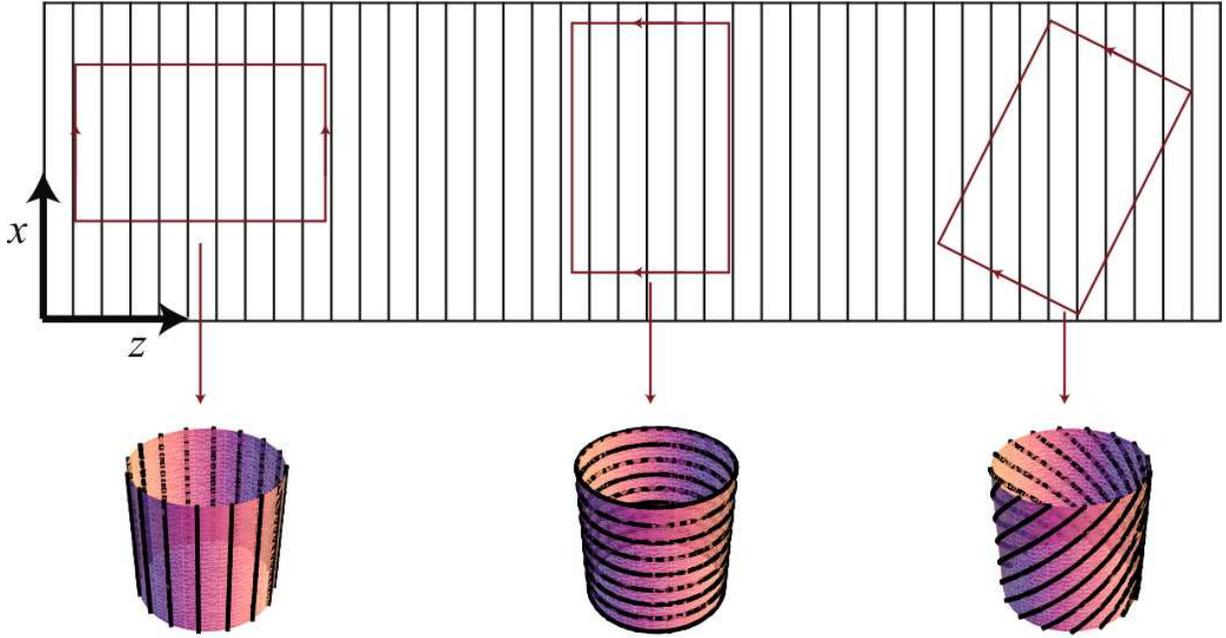}
\caption{From left to right: ``pin stripe,'' ``rugby stripe,'' and ``barber pole'' textures on a cylinder. These can be constructed from straight lines on a plane by identifying two sides of a rectangle (red). Though the layers are always geodesics, the layers of the left cylinder (``pinstripe'') are also straight. When $K=0$, the layers preferentially lie along the principle direction of zero curvature.}
\label{fig:cylinders}
\end{figure*}

In this paper, we study configurations of uniformly-spaced layers on a curved surface, elucidating and elaborating on the mechanisms of geometrical frustration and their effects on layer configurations. In addition, we explicitly consider the effect of the energetic cost of bending the layers to lie on the surface. This extrinsic bending effect leads to a purely geometrical ordering field for the layers. Though much effort has recently been directed toward this problem, particularly theoretical \cite{blanc01,santangelo07,xingPRL,xinglong} and numerical \cite{chantawansri07,li06,matthews03,varea99,pinna08}, this extrinsic curvature effect has received relatively little attention. In fact, bending energies can play an important role in ordering stripe patterns over longer length scales than would otherwise be possible \cite{santangelo07}. This observation suggests the use of surface curvature as a motif to control self-assembly of block copolymers \cite{connal08}.  Consider, for example, the problem of repulsive semi-flexible polymers on a cylinder.  In the fixed density ensemble, the polymers will attempt to adopt an equal spacing.  As shown in Fig. 1, there are many such structures ranging from ``pin stripes'' to ``rugby stripes'', as well as the continuous class of ``barber pole'' textures.  What do these have in common?  They all have equally spaced polymers and, from the point of view of {\sl intrinsic} geometry, they are all straight, that is, they are all geodesics.  This can be seen, for instance, by cutting open the cylinder and laying it flat -- all of 
these stripe textures map into straight lines on the flattened cylinder. However, these polymer-like lines are three-dimensional and their energy is a function of their embedded conformation in $\mathbb{R}^3$.  The ``pin stripe'' texture in Fig. \ref{fig:cylinders} is the only one for which the polymers have vanishing three-dimensional curvature and will thus be the ground state. 

\begin{figure*}
\includegraphics[width=0.9 \textwidth]{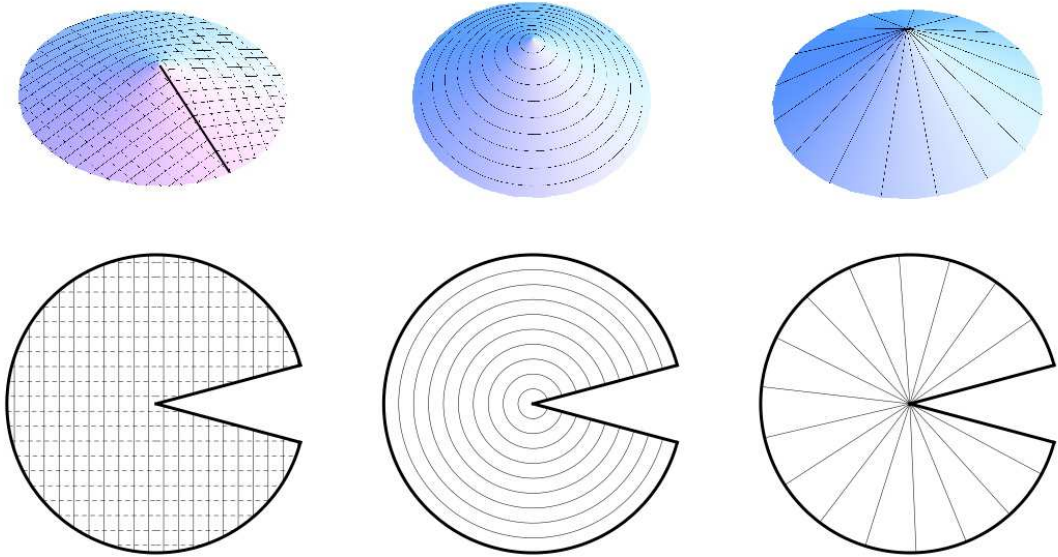}
\caption{From left to right: ``equally-spaced geodesics with sharp bend,'' ``equally-spaced lines with no bend singularities,'' and ``geodesics with vanishing extrinsic bending'' textures on a cone. These are all constructed by taking a pattern on a flat sheet, as could have been done for the cylindrical texture of Fig. \ref{fig:cylinders}.  A wedge angle is removed  as shown.  On the left cone the layers have equal spacing everywhere but for the line of dislocations along the seam and the layers have some normal curvature.  In the center, the layers are equally spaced but there is both geodesic and normal bending.  The right-hand cone has vanishing bending energy of all kinds, but the layer spacing is incorrect almost everywhere.  The dashed lines on the cone to the left indicate the layer normals.}
\label{fig:cones}
\end{figure*}

To this energetic accounting, we add a compression energy which favors equal-spacing between the lines and an intrinsic curvature energy within the local tangent plane.  We write the total free energy
schematically as:
\begin{eqnarray}\label{eq:schem}
F &=&[\hbox{compression energy}]\nonumber\\&&\quad + [\hbox{intrinsic bending of lines}] \nonumber\\&&\qquad+ [\hbox{extrinsic bending of lines}]
\end{eqnarray}
In Sec. \ref{sec:smectic} we will develop explicit, geometric expressions for these three contributions.  However, to frame our analysis, we pause to discuss 
the subtle frustration which arises from these competing terms.  The compression term measures the deviation of the layers from being equally spaced.   On a curved substrate it is still possible to define ``equal-spacing'': at each point on a given line we move along the geodesic tangent to the line's normal for a fixed interlayer distance (see, for instance, the right side of Fig. \ref{fig:gaussbonnet}).  This construction generates equally-spaced layers.  However, in contrast to a cylinder, on a curved substrate this
is not straightforward.  To see this, we consider lines on cones; the cone is flat everywhere but for its apex where it has a concentration of Gaussian curvature.
In Fig. \ref{fig:cones} we show three different smectic complexions on a cone.  Each is generated by first tracing a smectic pattern on a flat sheet.  Removing a wedge from the sheet allows us to construct a cone -- a surface with vanishing Gaussian curvature everywhere except at the point.  Geodesics on the cone become straight lines on the original surface and {\sl vice versa}.  On the left we see equally-spaced lines (solid) and their normal flows (dashed).  When put onto a cone, the geodesics remain equally spaced.  Since the intrinsic bending vanishes, by definition, for geodesics, this pattern has no contribution from the first two terms almost everywhere.  However, due to the Gaussian curvature conentrated at the tip of the cone, the global effect of the Gauss-Bonnet theorem (discussed below) is that a sharp, intrinsic bend must be introduced
into the layers along a seam of the cone to account for the angle deficit.  Moreover, though these lines are flat from the point of view of the surface, they are bent in $\mathbb{R}^3$ and thus
contribute to the \textit{extrinsic} curvature energy as well.  Since the directions along which the extrinsic curvature vanishes depends on how the surface
is bent into three-dimensions, this effect leads to an extrinsic field which can align the layers (see below).  
It is instructive to consider the other two configurations in Fig. \ref{fig:cones} as well.  In the center figure a set of concentric circles is wrapped onto a cone.  From
the construction we know that the resulting lines are equally-spaced and have no sharp bends.  The circles have \textit{both} intrinsic and extrinsic curvature; we can see this by noting that the normal to each circle points towards its center, in the same plane as
the circle.  However, the normal to the cone's surface does not lie in that plane and so
the circle's curvature has components both along the surface normal and perpendicular to it.
The former measures the extrinsic or normal curvature, the latter the intrinsic or geodesic curvature.  Finally, on the right, we show a set of lines which all converge to a point.  They are not equally-spaced on either the plane or the cone; by construction they are geodesics and for this particular embedding of the cone, they have vanishing extrinsic curvature.  The simple cone geometry highlights the geometric frustration embodied in simultaneously minimizing the three competing terms in the free energy.

In section II we introduce the geometric tools necessary to characterize stripes and develop the free energy that controls smectic order on curved surfaces.  In section III we present a number of examples that illustrate the effects of both intrinsic and extrinsic geometry on striped pattern formation.  Because of the many different limits that might be considered in this frustrated system, we focus primarily on zero-strain complexions where the lines are equally spaced (compression energy dominates).  This limit is closely related to the geometric optics limit of light propagation, as we will exploit. As we shall see, even with this constraint the competition between the remaining two terms is quite subtle. In section IV we discuss local mechanisms (related to extrinsic curvature energy) which lead to long range, two-dimensional order on curved surfaces, bypassing the standard Coleman-Mermin-Wagner conclusions about thermal fluctuations in flat space \cite{Coleman,MermWag}.  Finally, we summarize our results and discuss open questions.

\section{Layered structures on a curved substrate}

\subsection{Geometrical background}

Given the focus in general relativity and string theory on intrinsic geometric concepts such as conformal invariance, diffeomorphism invariance, and modular invariance, it is worth remembering that fascinating materials exist in three dimensions where not only intrinsic but also \textit{extrinsic} geometry plays a key role. 
With extrinsic curvature in mind, recall that an embedded surface ${\bf X}(u_1,u_2)\in\mathbb{R}^3$
has an induced metric $g_{ij}$ (first fundamental form), a unit surface normal $\bf N$, and a curvature tensor $L_{ij}$ (second fundamental form):
\begin{eqnarray}
g_{ij}&=&\partial_{i} {\bf X}\cdot\partial_j{\bf X}\nonumber\\\label{eq:geometrydefs}
{\bf N}&=&\frac{\partial_{1}{\bf X}\times\partial_{2}{\bf X}}{\left\vert\partial_{1}{\bf X}\times\partial_{2}{\bf X}\right\vert}\\
L_{ij}&=&-\partial_{i}{\bf N}\cdot\partial_{j}{\bf X} =  {\bf N}\cdot\partial_{i}\partial_{j}{\bf X}\nonumber
\end{eqnarray}
where the last equality follows from differentiating ${\bf N}\cdot \partial_{i}{\bf X}=0$.  
Note that our sign convention for $L_{ij}$ reflects the standard sign for the curvature of a three-dimensional space curve, {\sl i.e.} a circle will have a positive curvature when its normal is chosen to point \textit{inward}.  Here and throughout we refer the reader to \cite{diffgeom} for more technical details.

The two principle curvatures $\kappa_1$ and $\kappa_2$ are the eigenvalues of $g^{-1}L$, where the indices associated with $g_{i j}$ and $L_{i j}$ are suppressed to simplify the notation. The unit eigenvectors associated with the matrix $L$ are the principal directions ${\bf e}_1$ and ${\bf e}_2$.  On a general surface the curvatures and directions change from point to point.  With stripes, polymer strands, or smectic layers in mind, consider a curve embedded in the surface, ${\bf R}(s)={\bf X}\left(\sigma_1(s),\sigma_2(s)\right)$, with $s$ measuring the arclength.  The curve has unit tangent
${\bf T}(s)=d{\bf R}/ds$ with derivative $d{\bf T}/ds = \kappa(s){\bf N}_{\rm curve}(s)$ -- an equation which defines the unit normal to the curve, ${\bf N}_{\rm curve}(s)$, and the curvature, $\kappa(s)$.  Because ${\bf N}_{\rm curve}$ and the surface normal ${\bf N}$ defined above are not necessarily at right angles, we write
\begin{equation}
\frac{d{\bf T}}{ds} =\kappa{\bf N}_{\rm curve}= \kappa_n{\bf N}+\kappa_g \left({\bf N}\times{\bf T}\right),
\end{equation}
so that the change in ${\bf T}$ is decomposed into a vector along the surface normal, $\bf N$, and a vector in the surface, ${\bf N}\times{\bf T}$.  The coefficients $\kappa_n$ and $\kappa_g$ are the normal and geodesic curvature, respectively, and obey $\kappa^2=\kappa_g^2+\kappa_n^2$.  
A geodesic on a surface is a curve for which the change in the tangent vector has no components in the surface, that is $\kappa_g=0$. 
However, even a great circle on a sphere curves in three-dimensional space.  At any point on a curved surface, we can choose coordinates so that ${\bf e}_i=\partial_{i}{\bf X}$ are orthonormal so that $g_{ij}=\delta_{ij}$ and 
\begin{equation}
g^{-1}L = L= \kappa_1 {\bf e}_1{\bf e}_1^T + \kappa_2 {\bf e}_2{\bf e}_2^T,
\end{equation}
where $^T$ indicates the transpose \cite{diffgeom}.
It follows that the normal curvature can be further decomposed as
\begin{eqnarray}
\kappa_n &=& \mathbf{N} \cdot \frac{d \mathbf{T}}{ds} = -{\bf T}\cdot\frac{d{\bf N}}{ds}  = -{\bf T}\cdot\left({\bf T}\cdot\nabla\right){\bf N} = {\bf T}^TL{\bf T}\nonumber\\
&=&\kappa_1\cos^2\beta+ \kappa_2\sin^2\beta
\label{eq:kappan}
\end{eqnarray}
where ${\bf T}\cdot{\bf e}_1=\cos\beta$ gives the angle $\beta$ between the curve and one of the principle directions.  The last equality in the first line follows from the definition of $L$ in (\ref{eq:geometrydefs}) and from expanding $\bf T$ in terms of the two tangent vectors ${\bf e}_i$ \cite{diffgeom}.

On a radially symmetric surface (such as the Gaussian bump discussed below) with height function $h(r)$, ${\bf X}(r,\theta) = \left[r\cos\theta,r\sin\theta,h(r)\right]$.  The principal directions are, by symmetry, along $\mathbf{\hat{r}}$ and and along $\mathbf{\hat{\theta}}$.  It follows that if $\beta$ is the angle between the curve and the radial direction, then $\kappa_n=\kappa_r\cos^2\beta + \kappa_\theta\sin^2\beta$ where
\begin{eqnarray}
\kappa_r &=& \frac{\partial_r^2 h}{\left[1+\left(\partial_r h\right)^2 \right]^{3/2}}\nonumber\\
\kappa_\theta &=& \frac{\partial_r h}{r \left[1+\left(\partial_r h\right)^2\right]^{1/2}}\label{eq:kappa}
\end{eqnarray}
When $\kappa_r\kappa_\theta\le0$ (on the flanks of a Gaussian bump), there will always be an angle $\beta$ for which $\kappa_n$ vanishes, and so the extrinsic geometry and extrinsic bending energy of stripes on a surface set a natural, local direction for the polymers.  Now our discussion for smectic layers on cylinders (recall Fig. \ref{fig:cylinders}) can be made precise.  On a cylinder of radius $R$, $\kappa_r=0$ and $\kappa_\theta=1/R$ so $\kappa_n=0$ only when ${\bf T}\cdot{\bf e}_\theta=0$, which selects the preferred minimum energy configuration.

The cylindrical ground state is unambiguous because all of the textures in Fig. 1 are composed of lines drawn along geodesics.  Note that this {\sl extrinsic} effect is sensitive to an {\sl intrinsic} quantity, the angle $\beta$. On a smooth bump, with a nonzero, spatially varying Gaussian curvature, the problem is complicated by the intrinsic geometry of the curves defined by the layers.  Now we must also consider $\kappa_g$, the geodesic curvature.  As we will see in the following, the surface geometry may not only favor line configurations for which $\kappa_n\neq 0$ but, more
importantly, the surface geometry can prevent the stripes from achieving both equal spacing and vanishing $\kappa_g$.

\subsection{The Smectic Energy on Curved Substrates}\label{sec:smectic}

Lamellae, both on surfaces and in the plane, are conveniently represented by the level sets of a function $\Phi(u_1,u_2)=\Phi(\textbf{u}) = a n$, where $a$ is the equilibrium layer spacing and $n$ is an integer labeling the layer. The phase field $\Phi$ is related to the average mass density $\rho(\textbf{u})$ of the layers by $\rho(\textbf{u}) = \rho_0 + \rho_1 \cos (2 \pi \Phi/a)$. The gradient $\partial_i \Phi$ contains information both about the layer spacing, related to its magnitude, and the direction normal to the layers. 
We will use $\nabla$ to denote the covariant derivative on the surface, defined as $\nabla_i \Phi = \partial_i \Phi$ for a scalar function $\Phi$, and $\nabla_i v^j = \partial_i v^j + \Gamma^j_{i k} v^k$ for (contravariant) vectors \cite{diffgeom}. When unambiguous, we will also use $\cdot$ to indicate contraction: thus, $\textbf{v} \cdot \nabla \Phi = v^i \partial_i \Phi$. Note that $\mathbf{v} \cdot \nabla \Phi$ gives the change in $\Phi(u)$ when one moves in an arbitrary direction $\mathbf{v}$ on the surface. Upon specializing to the case where $\mathbf{v}=\nabla \Phi$, the resulting change is $| \nabla \Phi|^2 = \nabla \Phi \cdot \nabla \Phi = g^{i j} \partial_i \Phi \partial_j \Phi$.
Hence, equal layer separation implies the condition $\vert\nabla\Phi\vert^2 = 1$.  The strain, $e$, measures the deviation from equal spacing.  While there are many possible forms for $e$, depending on microscopic details, we only require that $e$ vanish when $\vert\nabla\Phi\vert^2=1$ and, in the small deformation limit where the Eulerian displacement is $u=z-\Phi$, that $e\rightarrow \partial_z u$ where $z$ is the local direction of the layer normal.  A suitable form is 
\begin{equation}\label{eq:straindef}
e =  \frac{1 - |\nabla \Phi|^2}{2}.
\end{equation}
The energy cost of small deformations from equal spacing will be an expansion in powers of $e$, with the lowest order term being $e^2$.   If the layers are equally spaced, then $1=\vert\nabla\Phi\vert^2$ and, upon differentiation, we find that $0=\left(\nabla\Phi\cdot\nabla\right)\nabla\Phi = \left({\bf n}\cdot\nabla\right){\bf n}$ where $n^i= g^{ij}\partial_j\Phi/\vert\nabla\Phi\vert$ is normal (within the tangent plane) to the curve used to define a layer.  
The condition $\left({\bf n}\cdot\nabla\right){\bf n}=0$ is precisely the geodesic equation for the layer normal.  We see then, independent of the form of the strain, when $e=0$ (thus, minimizing the strain energy) the layers are spaced evenly along geodesic curves.  
If we consider only zero strain solutions, this condition reduces the problem of finding smectic textures to a \textit{first-order} evolution equation. Thus, the normals lie on ``straight lines'' on the surface (\textit{i.e.} geodesics); layer curvature leads to their convergence into singularities.

Indeed, if $\xi$ measures the distance between two geodesics along a third, perpendicular geodesic, the best local approximation to parallel trajectories leads to \cite{diffgeom}
\begin{equation}\label{eq:divergence}
\frac{d^2 \xi}{ds^2} = - K \xi,
\end{equation}
where $K$ is the Gaussian curvature.
This geodesic deviation equation naturally reminds us of geometric optics \cite{landau-elec},
in which the layer normals act as rays and the Gaussian curvature, $K(u_1,u_2)$ as a variable index of refraction. In regions of negative curvature the normals diverge as through a diverging lens; in regions of positive curvature they converge.  Converging patterns of rays form caustics; in the language of ordered lamella, caustics are curvature singularities in the lines.
On a curved surface, a set of geodesics normals initially perpendicular to some layer may cross a finite distance away, leading to a cusp-like  boundary where the geodesics finally converge.  This system thus provides a low-dimensional analog of gravitational lensing \cite{Lensing,VJK}.  Cusps are regions of high bending energy and thus, in the true ground state, they will be smoothed out at the cost of compression energy.

We are now in a position to formulate the complete free energy:
\begin{equation}\label{eq:energy}
F = \frac{B}{2} \int dA~\left[ \frac{\left(1 - |\nabla \Phi|^2 \right)^2}{4} + \lambda_g^2 \kappa_g^2 + \lambda_n^2\kappa_n^2 \right],
\end{equation}
where $B$ is the bulk modulus and the couplings $\lambda_{g}$ and $\lambda_{n}$ are length scales, typically on the order of a column diameter, which measure the relative importance of bending and compression energies.  In principle $\lambda_g\neq \lambda_n$: the $\lambda_n\rightarrow 0$ limit is the ``generally covariant'' limit where extrinsic effects are irrelevant.    On the other hand, if we were trying to decorate the surface with semi-flexible polymers with persistence length $L_P$, we would have $\lambda_g^2=\lambda_n^2=k_BT L_P h/(Ba^2)$, where $h$ is the thickness of the polymer layer normal to the surface and $a$ is the average spacing between the polymers \cite{cline}. 
In general, nonzero  $\lambda_g$ and $\lambda_n$ will lead to corrections to the simple picture of caustic singularities sketched above.  

To better understand Eq. (\ref{eq:energy}), consider the three different decorations of the cone shown in Fig. 2, assuming finite cones of radius $R$.  On the left, while there is some extrinsic layer bending leading to an energy on the order of $B\lambda_n^2\ln R$, the dominant energetic cost arises from the bend wall on the seam: there the equal-spacing condition will breakdown in order to smooth out the sharp kink.  Equivalently, we can view this as a row of dislocations as arises in low-angle grain boundaries with elastic core energy per unit length scaling as $Ba^2$.  In either case, we see that the energy
of the grain seam scales as $F_{left} \sim BaR$.  Upon focussing on the central cone, we see that the smectic complexion has energetic contributions from both
bending moduli and $F_{\rm center}\sim B(\lambda_g^2 +\lambda_n^2)\ln R$.  Consider finally the cone on the right; the texture there has no bending energy at all, but a very large has compression energy, $F_{\rm right} \sim BR^2$.  From this rough analysis, we would expect that away from regions of large Gaussian curvature (in this case the conical singularity), the ``bulls-eye'' pattern of the center cone of Fig. 2 will be the dominant line texture.  However, depending on the various elastic constants this can break down for finite regions around the the peak.  In the following we will consider these effects, in particular in Section IV.  

For further insight into Eq. (\ref{eq:energy}), it is instructive to consider this free energy in the context of weakly deformed two-dimensional smectics coating a cylindrical surface with a flat metric and no singularitiy. We introduce the usual one-dimensional displacement field $w(u_1,u_2)$ and consider the simple case of a cylinder discussed in the introduction. In the coordinate system of Fig. \ref{fig:gaussbonnet}, with $(u_1,u_2) = (x,z)\equiv \mathbf{r}$, we take as our level set function
\begin{equation}
\Phi(x,z) = \mathbf{\hat{z}} \cdot \mathbf{r} - w(\mathbf{r}).
\end{equation}
Consider the limit of gentle, slowly-varying undulations superimposed on the texture shown at the bottom left of Fig. \ref{fig:gaussbonnet}. For a cylinder of radius $R$, it is straightforward to show that the free energy then becomes
\begin{equation}\label{eq:nelson2}
F = \frac{B}{2} \int dA~\left[ \left(\partial_z w \right)^2 + \lambda_g^2 \left(\partial_x^2 w\right)^2 + \left(\lambda_n/R\right)^2 \sin^4 \left(\partial_x w\right) \right],
\end{equation}
where $\partial_x w(\mathbf{r}) \approx \beta$ is the local tilt angle that the stripes make with respect to the $x$-axis. The last term in Eq. (\ref{eq:nelson2}) follows from Eq. (\ref{eq:kappan}), where we choose axes of principle curvature such that $\kappa_1=0$ and $\kappa_2 = 1/R$. In addition to the usual terms describing a rotationally-invariant 2d smectic \cite{nelsonbook}, there is now a weak ordering field proportional to $\sin^4 \beta$. The field is weak in two senses -- its strength is proportional to $1/R^2$ and hence vanishes in the limit of a cylinder of infinite radius. Furthermore, it goes like $\beta^4$ for small tilt angle deviations from the preferred direction. Nevertheless, this field is enough to break the symmetry and give rise to a preferred direction for the lines. As discussed above, this field arises from the extrinsic curvature tensor and will vary spatially in both magnitude and direction on more general curved surfaces. In the remainder of this paper, we first neglect this weak field (for a Gaussian bump of height $h_0$ and size $R_0$, $h/R_0^2$ replaces $1/R$) and focus on satisfying the constraint of equal layer spacing imposed by the first term of Eq. (\ref{eq:energy}), with a simplifying boundary condition at infinity. We then discuss effects due to this extrinsic ordering term with free boundary conditions at infinity.

An anisotropic membrane would have two bending moduli, one for bends along a ``hard'' direction, the other for bends along the ``soft'' direction.  If the anisotropy had a vector character, we could also include a cross term, allowed by symmetry.
Defining $\textbf{t}$ as the unit vector lying along the layers, an alternate form for the normal curvature is $\kappa_n = t^i t^j L_{i j}$, where $L_{i j}$ is the surface curvature tensor given in equation (\ref{eq:geometrydefs}). The other two allowed terms are  $n^i n^j L_{i j}$ and $n^i t^j L_{i j}$, respectively. 
The magnitude of the moduli for these additional terms depend on the molecular details of the smectic layers. These elastic constants are different, in principle, from the intrinsic bending modulus $B\lambda_g^2$ in (\ref{eq:energy}).  For concreteness, we will focus our discussion on a monolayer of block copolymer cylinders lying on the surface.  
The columnar phase of neat (\textit{i.e.} monodisperse with no solvent) block copolymers, though bearing resemblance to the columnar hexagonal phase of liquid crystalline polymers, is actually an incompressible three-dimensional elastic medium.  Strong-segregation calculations \cite{GK} suggest that the columns can be viewed as semi-flexible rods.  In that case ``bending along the columns'' is the ``hard'' direction, being more energetically costly than ``bending perpendicular to the columns'' for a few-layer coating of a curved substrate.  Roughly speaking, the columnar phase is similar to a corrugated sheet with the columns corresponding to the corrugations.  However,
as the diblock film grows, bending along the columns leads to deformations that are independent of the thickness since lamella can slide past each other with no cost.  On the other hand, bending perpendicular to the columns requires large amounts of crystalline strain.  For a thin film composed of only a few layers, the bending energy we consider here should dominate.  However, how the introduction of low-angle grain boundaries and surface energies affects these calculations is an open question \cite{GK}.  To keep our analysis
from becoming highly ramified, the only non-vanishing extrinsic elastic modulus will be
bending along the column tangents.  

\subsection{Mechanisms of geometric frustration}

With the spate of recent work on both crystalline order and nematic order on curved substrates, it is valuable to compare and contrast these systems with the smectic -- a phase with one-dimensional crystalline order that lives \textit{between} the crystal and the nematic.  We shall see that the smectic presents issues all its own and affords a fresh arena for the interplay between geometry and soft materials.  

A nematic liquid crystal on a surface is described by a unit vector $\bf n$ which lives in the tangent plane of the surface.  The standard three-dimensional Frank free energy \cite{deGennes}
\begin{eqnarray}
F[{\bf n}] &=& \frac{1}{2}\int d^3\!x\,\left\{K_1\left(\nabla\cdot{\bf n}\right)^2 + K_2\left[{\bf n}\cdot\left(\nabla\times{\bf n}\right)\right]^2\right.\nonumber\\
&&\qquad\left. + K_3\left[\left({\bf n}\cdot\nabla\right){\bf n}\right]^2\right\}
\end{eqnarray} 
is modified in two ways:  First, ${\bf n}\cdot\left(\nabla\times{\bf n}\right)=0$ when ${\bf n}$ lies on a smooth, two-dimensional surface and depends only on the two surface coordinates ${\bf u}=\{ u_1,u_2 \}$.
The two-dimensional nematic free energy then reads
\begin{equation}\label{eq:nematic}
F_{nem} = \frac{K_1}{2} \int dA \, \left(\nabla \cdot \textbf{n}\right)^2 + \frac{K_3}{2} \int dA\,\left[ \left(\textbf{n} \cdot \nabla \right) \textbf{n}\right]^2,
\end{equation}
where $dA=d^{2}u \sqrt{g}$ and $g$ is the determinant of the metric
tensor $g_{ij}$. The first term in Eq. (\ref{eq:nematic}) penalizes director splay, whereas the second term penalizes director bend. When $K_1=K_3 \equiv K_F$ the free energy
in Eq. (\ref{eq:nematic}) is isotropic and can be cast in a form that naturally lends geometric insight.  

Consider a local angle field $\beta({\bf u})$,
corresponding to the angle between ${\bf n}({\bf u})$ and an
arbitrary, orthonormal local reference frame whose basis vectors we label
${\bf e}_{i}({\bf u})$ with $i=1,2$. In the one Frank constant approximation, the free
energy in Eq. (\ref{eq:nematic}) can be recast in the form
\begin{equation}
F = \frac{1}{2}K_F \int dA \!\!\!\!\! \quad g^{ij}
(\partial_{i}\theta - \Omega_{i})(\partial_{j}\beta - \Omega_{j})\!\!\!\!
\quad , \label{eq:patic-ener}
\end{equation}
where $\Omega_{i}({\bf u})$ is a connection that plays the role of the Christoffel symbols by compensating for
the rotation of the 2D basis vectors ${\bf e}_{i}({\bf u})$
in direction $i$ and $\nabla\times\mathbf{\Omega}=K$ \cite{kamien}.  Since the curl of
$\Omega_{i}({\bf u})$ is equal to the Gaussian curvature $K({\bf u})$,
the nematic energy cannot vanish
on a surface with non-zero Gaussian curvature. Note that
$\Omega_{i}({\bf u})$ is, in general, a non-conservative field, so we cannot minimize (\ref{eq:patic-ener}) by setting $\partial_i \beta$ equal to $\Omega_i$ everywhere on the surface.
This property is a manifestation of a more general mechanism, commonly referred to
as geometric frustration, to indicate situations where the molecular arrangement favored by local interactions cannot be extended globally.

As the Gaussian curvature of the substrate increases, defects are generated in the ground state to lower the energy cost of geometric
frustration. Their energetics is analogous to Coulomb particles interacting with a smeared out electrostatic charge given by the Gaussian curvature.
This nontrivial result can be rationalized by examining the free energy in Eq. (\ref{eq:patic-ener}) and noticing that the connection $\Omega_{i}({\bf u})$ and the Gaussian curvature $K({\bf u})$ are analogous (in \textit{two} dimensional electromagnetism) to a frozen vector potential and the magnetic field respectively. The topological defects, {\sl i.e.} disclinations, behave as monopoles in the dynamical field $\beta({\bf u})$ whose interaction with the geometry of the surface
is mediated by the geometric gauge field $\Omega_{i}({\bf u})$. An additional coupling between defects and the metric arises from the metric factors that appear, for example, in the surface element $dS=d^{2}u\sqrt{g}$ independently of the connection \cite{ViteAri}.

The physics of geometric frustration is at work also in the more complicated setting of curved space crystallography  \cite{Bow00,Giomi07,vitelliPNAS}. Now the orientational order of the nematic is supplemented by translational degrees of freedom. For gently deformed surfaces, the crystalline energy
can be expressed in terms of the Lam\'e coefficients $\mu$ and $\lambda$ \cite{Lan99}
\begin{equation}
 F= \int\! dA \!\!\!\!\! \quad \left( \mu \!\!\!\!\! \quad u_{ij} ^{2}(\vec{x}) + \frac{\lambda}{2} \!\!\!\!\! \quad u_{kk} ^2(\vec{x})   \right) \!\!\!\! \quad ,
\label{eq:lame}
\end{equation}
where $\vec{x}=\{x,y\}$ are cartesian coordinates in the plane and $u_{ij} (\vec{x})=\frac{1}{2}\left[\partial_{i} u_j (\vec{x})+ \partial_{j} u_i (\vec{x})+ A_{ij} (\vec{x})\right]$ is the strain tensor. Compared to its flat space counterpart, the strain tensor has an additional term $A_{ij}(\vec{x})=\partial_{i}h(\vec{x}) \partial_{j}h(\vec{x})$ that couples the gradient of  the displacement field $u_{i}(\vec{x})$ to the gradient of the surface height function $h(\vec{x})$. The field $A_{ij}(\vec{x})$ is a tensor version of the connection $\Omega_{i}$ introduced above to describe orientational order on curved surfaces. Indeed the curl of the tensor field $A_{ij}(\vec{x})$ is equal to the Gaussian curvature of the surface $K(\vec{x}) = -\epsilon_{il}\epsilon_{jk}\partial_{l}\partial_{k} \partial_{i}h(\vec{x}) \partial_{j}h(\vec{x})$ where $\epsilon_{ij}$ is the antisymmetric unit tensor ($\epsilon_{xy}=-\epsilon_{yx}=1$) \cite{Sac84,witten}. By the same reasoning as before, the integrand of Eq. (\ref{eq:lame}) and hence the ground state energy cannot be made to vanish. This is the mathematical mechanism by which geometric frustration enters the physics of curved crystals. It can be grasped more intuitively by recalling that bending a plate into a surface of non vanishing Gaussian curvature necessarily causes it to stretch \cite{Lan99}.    
 
Smectic liquid crystals, on the other hand, can maintain uniform layer spacing (and hence achieve zero strain) even in the presence of Gaussian curvature. The Gaussian curvature, nevertheless, couples to the curvature of the layers. To see how this coupling appears, we start by noticing that the bend coupling proportional to $K_3$ in equation (\ref{eq:nematic}) resembles the geodesic equation for curves tangent to the director. We can establish an intuition for smectic patterns, therefore, by studying nematics with very large $K_3$, and identifying the nematic order parameter as the layer normal. Nematics on spheres are required to have a net $+2$ topological charge, which tend to break up into four $+1/2$ disclinations. When $K_3 = K_1$, the disclinations lie on the corners of a tetrahedron \cite{nelson02,vitelli06}; when $K_3$ is large, however, the $+1/2$ disclinations 
lie on an equator \cite{shin08} and the local texture is a lines of longitude structure.  At the other extreme, when $K_1$ is large the director takes on a lines of latitude texture.

\begin{figure}
\includegraphics[width=0.45 \textwidth]{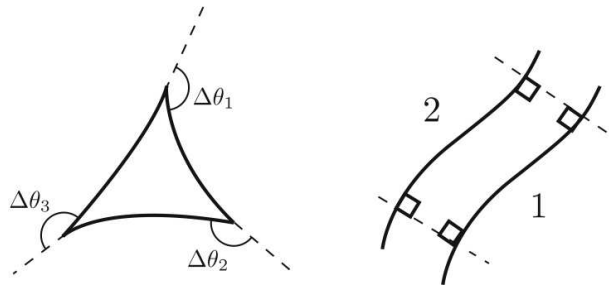}
\caption{Left: Turning angles $\Delta \theta_i$ defined on a triangle. Right: A square with analogous turning angles formed from two uniformly-spaced layers (solid) and two normals (dashed). The normals are geodesics so $\kappa_g = 0$ along them.}
\label{fig:gaussbonnet}
\end{figure}

Since there is no difficulty to finding geodesics in any direction at any point on a curved surface, there is no local obstruction to constructing equally-spaced layers on any surface. However, doing so while simultaneously finding layers which are also geodesics is impossible. This can be seen by constructing a rectangle with two opposite sides given by adjacent layers and the remaining two sides given by geodesics normal to both layers (see Fig. \ref{fig:gaussbonnet}). Upon applying the Gauss-Bonnet theorem \cite{diffgeom} to this contour, we obtain
\begin{equation}\label{GB}
\int dA\,K = 2 \pi - \sum_i \Delta \theta_i - \oint ds\,\kappa_g,
\end{equation}
where $\Delta \theta_i = \pi/2$ is the turning angle at each corner of the rectangle, we find

\begin{equation}\label{eq:GB2}
\int dA\,K =\int_2 ds~\kappa_g - \int_1 ds\,\kappa_g.
\end{equation}
The integral on the left-hand side is over the area of the square while the integrals on the right-hand side are over the two adjacent layers. Since $\int dA\,K \ne 0$ in general, it is not possible for $\kappa_g = 0$ on both layers.

Despite our local ability to set $e=0$ in equations (\ref{eq:straindef}) and (\ref{eq:energy}), there may be global obstructions to finding low energy configurations with vanishing compression strain.  To see this, we first recast (\ref{GB}) locally.  Upon using the local coordinates $(x,y)$ and introducing the surface metric $g_{ij}$, Eq. (\ref{eq:GB2}) becomes
\begin{equation}
\int dx dy \sqrt{g} K = \frac{d}{dn} \int_1^2 dn  \int ds\; \kappa_g
\end{equation}
where $d/dn$ is the derivative along the normal direction and we have used the fundamental theorem of calculus.  After rewriting the integral on the right in the local coordinates, we come to the local relation
\begin{equation}\label{eq:GB3}
\sqrt{g} K = {\partial_n}\left(\sqrt{g} \kappa_g\right)
\end{equation}
Recall that the geodesic curvature is the fractional rate of change of the length of an arc as it is moved perpendicular to itself and that positive curvature implies that normal evolution shrinks the curve \cite{diffgeom}.  It follows that  $\kappa_g = - g^{-1/2}\partial_n \sqrt{g}$ where we define the normal direction to be along the curve's (inward) pointing normal.

Using Eq. (\ref{eq:GB3}), we find $\sqrt{g} K = (\partial_n \sqrt{g}) \kappa_g + \sqrt{g} \partial_n \kappa_g$. Dividing through by $\sqrt{g}$, we find
\begin{equation}\label{eq:evolution}
\partial_n \kappa_g = \kappa_g^2 + K.
\end{equation}
This evolution equation encapsulates the geometric frustration implicit in the Gauss-Bonnet theorem. Indeed, following the same reasoning, Eq. (\ref{eq:divergence}) is also a local form of Gauss-Bonnet;  we may take an arc of length $\xi$ connecting the two diverging (converging) geodesics at $s$.  Then the curvature of those arcs is $\kappa_g=-\xi^{-1}(d\xi/ds)$, and it follows that $-K\xi=d^2\xi/ds^2=-d\kappa_g/ds \xi -\kappa_g(d\xi/ds)$, which is identical to (\ref{eq:evolution}).

\section{Smectic Scattering from Curved Surfaces}

In this section, we review and generalize our prior results on smectic textures on a simple Gaussian bump \cite{santangelo07} to more complex geometries.

\subsection{The Gaussian Bump}
Consider smectic order on a curved substrate described in the Monge representation by a height function $h(x,y)$; the Gaussian curvature is then given by \cite{diffgeom}
\begin{equation}
K = \frac{\partial_x^2 h \partial_y^2 h - \left(\partial_x \partial_y h\right)^2}{\left[1+\left(\partial_x h\right)^2 + \left(\partial_y h\right)^2\right]}.
\end{equation}
If the whole surface can be described by $h(x,y)$ then it is topologically equivalent to the plane and, by the Gauss-Bonnet theorem, $\int dA\,K = 0$.  We start with the Gaussian bump used to study nematic order in Ref. \cite{vitelli06}, $h(r) = h_0 e^{-r^2/(2 R^2)}$, with $r^2=x^2+y^2$. $K$ is positive near $r=0$ and negative for large $r$. At an intermediate radius $r = R$, $K$ vanishes.
A particularly simple smectic configuration results from choosing radial geodesics as the layer normals, in which case we would find uniformly-spaced, azimuthal layers, a disclination at the top of the bump, and a power law decay of $\kappa_g$ away from the center.  Is it possible, however, to generate a configuration \textit{free} of topological defects on a bump? 

\begin{figure}
\includegraphics[width=0.45 \textwidth]{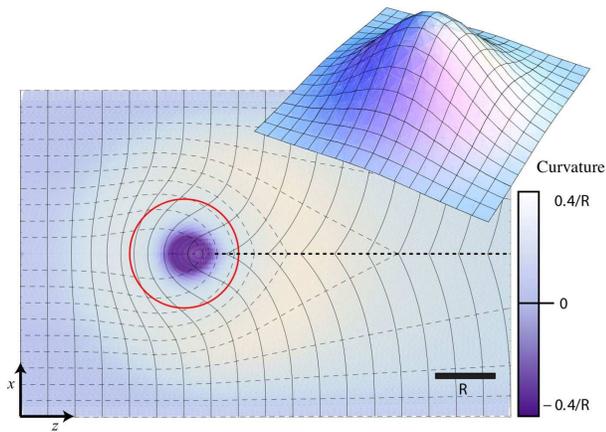}
\caption{Layers (solid lines), normals (dashed lines) and geodesic curvature (color) for a bump with aspect ratio $h_0/R = 3$ projected onto the $xy-$plane. The circle of zero Gaussian curvature is depicted in red. A scale bar of length $R$ has been provided. With the constraint of equally-spaced layers along the normals, the curvature induces the formation of a grain boundary (heavy dashed line) that extends infinitely far to the right of the bump center. The apparent unequal layer spacings in the figure are an artifact of the projection. The inset shows the bump, in perspective, overlaid with a square grid. Note that the grid lines are not equally-spaced in this case.}
\label{fig:layers}
\end{figure}

Suppose we start with straight layers at $x=-\infty$, so that the layer function $\Phi(x,y)$ defined in Sec. \ref{sec:smectic} obeys $\Phi \rightarrow x$ as $x \rightarrow -\infty$.  This boundary condition would describe an experimental setup where the layers of a diblock columnar phase grow along a temperature gradient parallel to $\hat x$.  We expect layer by layer growth nucleated from a boundary at large negative $x$.   The geometry of the substrate leads to the formation of singularities, but rather different than the isolated disclination with concentric circular layers centered on the bump discussed above (see Fig. \ref{fig:layers}).   We call these singularities caustics because, just as in geometrical optics, these are places where many initially parallel light rays or, in our case, geodesics converge.   At these locations the value of $\Phi$ is well-defined but $\nabla\Phi$ is discontinuous.  Though it might be tempting to call these cusps ``{\sl defects}'', they are not.  Recall that near a dislocation the smectic order parameter vanishes and the phase field $\Phi$ takes on {\sl all} values around the defect.  Similarly, in the vicinity of a disclination, the nematic order vanishes and the layer normal takes on all directions at the defect.  In contrast, the cusp singularities in Fig. \ref{fig:layers} have definite values of $\Phi$ and $\nabla\Phi$ does not wind through all possible directions around these singularities. Thus, although there is a discontinuity in $\nabla\Phi$, it is not of the same nature as that of a disclination.  In the analogy to optics, we would say that a dislocation is a place of vanishing amplitude, while a caustic is a location of very high, if not infinite, amplitude \cite{Nyebook}.  Figure \ref{fig:layers} shows a birds-eye view of a Gaussian bump coated with stripes, with the geodesics as dashed lines and the layers themselves as solid lines.  The red circle indicates the locus of points for which $K=0$ and we have shaded the regions according to the magnitude of 
$\kappa_g$.
We could also have predicted these cusps by determining where the curvature of the smectic layers diverges along the geodesics defined by the layer normal.  To do this, we integrate
 (\ref{eq:evolution}) along the normal geodesic passing over the top of the bump. We find that $\kappa_g$ diverges a finite distance past the center of the bump, shown in Fig. \ref{fig:layers}; this divergence indicates the onset of the infinite curvature cusps in the layer lines.

From a more global perspective, the cusp angle is a measure of the integrated Gaussian curvature. 
To see this, we again use the Gauss-Bonnet theorem to describe evolution of the cusp angle as a function of distance from the center of the bump. The key is to define a geodesic triangle with one edge along the $x$-axis, another edge along the layer parallel to the $y$-axis at $x \rightarrow -\infty$ and a third edge along a normal curve, which is a geodesic by construction. The resulting geodesic triangle has two exterior (or interior) $\pi/2$ angles. Denote the remaining internal angle by $\alpha$. The corresponding external angle, $\pi-\alpha$, is constrained by the Gauss-Bonnet theorem, which leads to
\begin{equation}
 \int_T dA\,K = 2\pi - \left[\frac{\pi}{2} +\frac{\pi}{2} + \left(\pi-\alpha\right)\right] =\alpha
\end{equation}
where the integral is taken over the inside of the triangle. Notice that, as the cusp location $x$ becomes large and positive (see Fig. \ref{fig:layers}), $\alpha(x) \rightarrow 0$ since the triangle incorporates an increasing amount of area in the entire upper half-plane. Because the integrated Gaussian curvature is zero in the upper half-plane, the angle $\alpha$ necessarily decreases with increasing $x$. Though the grain boundary persists infinitely far to the right of the center of the bump, the cusp angle asymptotically vanishes.  In the case of a Gaussian bump, the angle will fall off as $e^{-(r/R)^2}$, though the details will vary for other surfaces.  Similar to geometric optics, the formation of these cusps should not depend sensitively on the exact geometry of the substrate; rather it is a function of the topology as characterized by the intrinsic curvature.

\subsection{Smectic optometry}
\begin{figure}
\includegraphics[width=0.5 \textwidth]{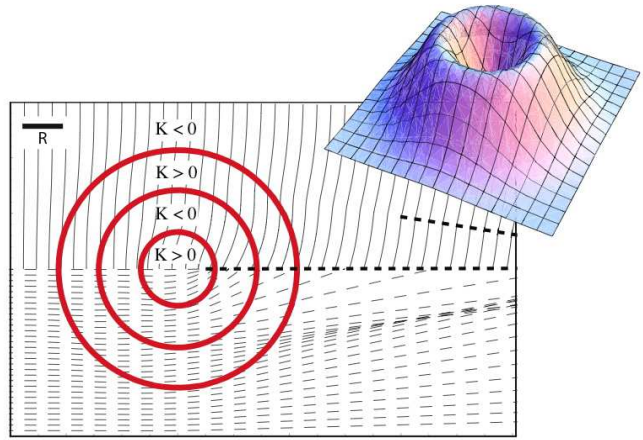}
\caption{ A converging lens: Layers (solid lines, only upper half-plane shown) and normals (dashed lines, only lower half-plane shown) for a bump with $h =  h_0 [(x^2 + y^2)/R] \exp [-(r/R)^2/2)$ ($h_0/R = 1$). A scale bar of length $R$ has been provided. All quantities exhibit mirror symmetry about the horizontal midline. Boundaries between regions of Gaussian curvature with different signs are delineated in red. Grain boundaries, the analogues of caustics in geometrical optics, are shown as bold dashed lines. Note the focusing of the normal lines onto the grain boundary in the lower half-plane. All lengths are measured in units of the bump width $R$. 
The surface is shown in perspective in the inset.}
\label{fig:x2plusy2}
\end{figure}

We have considered other substrates, all topologically equivalent to the plane so that again, the integrated Gaussian curvature vanishes, $\int K\,dA=0$.  
In Fig. \ref{fig:x2plusy2} we depict both the geodesics and layers generated from the same boundary condition as $x\rightarrow -\infty$ that we considered in the last section; now, however, we have chosen a more complex axisymmetric bump with multiple regions of positive and negative $K$.  Note that geodesics that do not go through the central region of the bump, outlined by the outermost red 
circle, will generate a set of cusps at large $x$ along the midline similar to those in Fig. \ref{fig:layers}.  
However, we now find that there are additional caustic lines that form as a result of the focussing from the inverted dimple of the bump.  These ``fold'' caustic lines, in this case, converge toward the $x$-axis and eventually end there.  To see this, we adapt the discussion in the last section: geodesics that avoid the central region entirely lead to larger-angle cusps on the $x$-axis than those that go through the same region.  At large $x$, the pattern must be that of the simple bump in Fig. \ref{fig:layers} and thus there will be no auxiliary caustics at large $x$ and $y$.  It follows that any extra caustics must converge to the $x$-axis.  We can see this in another way by considering the focussing equation (\ref{eq:divergence}).  Pairs of geodesic normal lines that remain in the outer region always diverge, because $K<0$ \cite{footnote1}.  However, a pair of geodesics, one of which remains in the outer region and which enters the inner region will still diverge less slowly and, indeed, pairs which both travel through the annulus with $K>0$ will {\sl converge}.  Hemmed in by the pairs that are always diverging, we see that these additional cusps must form and that the ``outer'' geodesics overtake the ``inner'' geodesics, leading to two extra cusp line grain boundaries, one of which is shown in the lower half of Fig. \ref{fig:x2plusy2}.

\begin{figure}
\includegraphics[width=0.5 \textwidth]{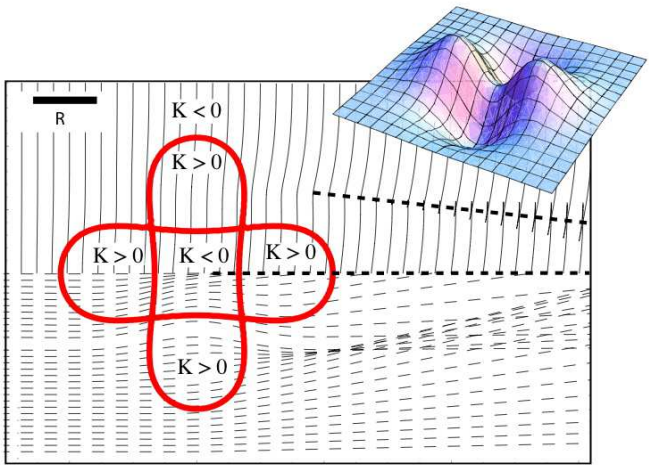}
\caption{ Layers (solid lines, only upper half-plane shown) and normals (dashed lines, only lower half-plane shown) for a saddle-bump with $h(x,y) = 0.6 [(x^2 - y^2)/R] \exp [-(r/R)^2/2]$. A scale bar of length $R$ has been provided. Boundaries between regions of Gaussian curvature with different signs are delineated in red. Grain boundaries are shown as heavy dashed lines.  
The surface is shown in perspective in the inset.}
\label{fig:x2minusy2}
\end{figure}

\begin{figure}
\includegraphics[width=0.5 \textwidth]{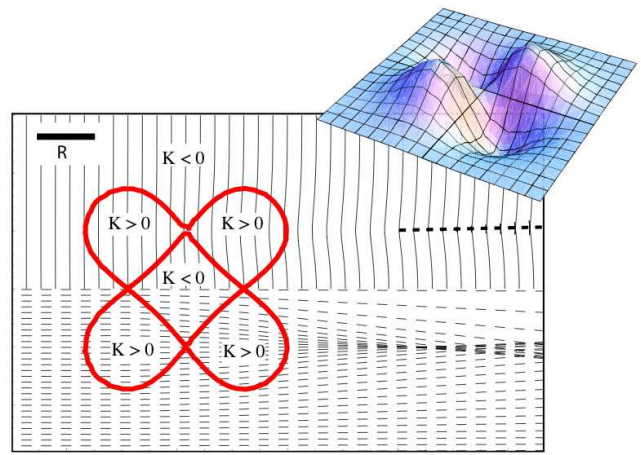}
\caption{A diverging lens: Layers (solid lines, only upper half-plane shown) and normals (dashed lines, only lower half-plane shown) for a bump with $h(x,y) = 0.55 (x y/R) \exp [-(r/R)^2/2]$. A scale bar of length $R$ has been provided. Boundaries between regions of Gaussian curvature with different signs are delineated in red. Grain boundaries are shown as heavy dashed lines. 
The surface itself is shown in the inset.
}
\label{fig:xy}
\end{figure}

\begin{figure*}
\includegraphics[width= \textwidth]{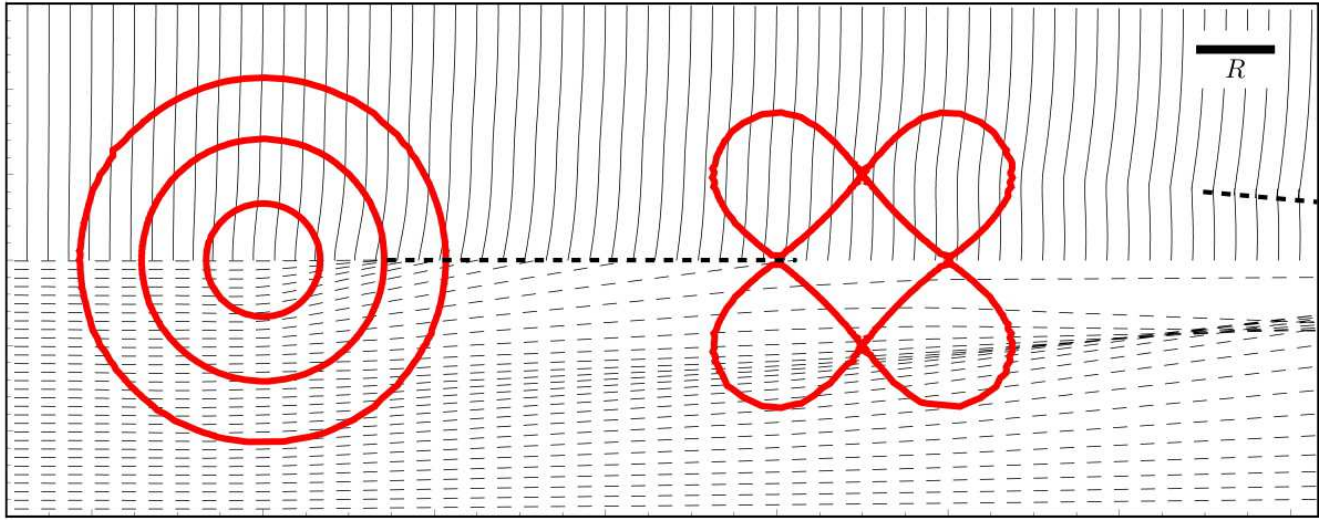}
\caption{Smectic optometry: a compound ``lens'' built from the bumps described in Fig. \ref{fig:x2plusy2} and \ref{fig:xy}. Layers are again denoted by solid lines and layer normals by (light) dashed lines. A scale bar of length $R$ has been provided. The grain boundary that appears due to the convergence of the left-hand bump ends at the diverging bump on the right. A second pair of caustics/grain boundaries appear to the right of this bump. Only one is shown, as the heavy dashed line.}
\label{fig:compound}
\end{figure*}

Just as microscopists design and choose multiple lenses to manipulate an image, one might hope to engineer a surface whose geometry leads specific smectic textures. To further our intuition, we have studied the non-axisymmetric saddle-bump shown in Fig. \ref{fig:x2minusy2}, with the polynomial $x^2-y^2$ multiplying a Gaussian envelope.  As shown in the inset, this surface has a saddle in the center, surrounded by four lobes of positive curvature.  
Upon choosing the same boundary conditions as $x \rightarrow -\infty$ as before, we find a line of cusps along the $x$-axis and two extra off-axis caustic lines that disappear as they approach $y=0$.  In order to find an example for which the auxiliary lines {\sl diverge} (a diverging instead of converging lens), we consider, for instance, the same saddle-bump structure, but rotated $45^\circ$ so that the geodesics that start near the $x$-axis minimize their transit across regions of positive curvature.  These geodesics are always diverging, picking up only a small amount of $K>0$ as they exit the central saddle region.  As a result, there are no cusps on the $x$-axis -- the layers are perfectly smooth since the geodesics associated with the layer normals are not focused to this line.  As shown in Fig. \ref{fig:xy}, one can see that the cusps in the layers form symmetrically, off-axis and that, for the bump shown, they bend away from the line $y=0$; it would be impossible for them to converge since the layers are regular on the $x$-axis.  It seems likely that, by adjusting the saddle-bump parameters, the lines of cusps could be made parallel to $\mathbf{\hat x}$.
These simple, high-symmetry cases provide intuition for how layer focussing could be used to construct desired patterns far from the bump.  It is amusing to consider other boundary conditions (for instance, a circular or parabolic arc at infinity incident on a bump with little or no symmetry.  Certainly the ``optical'' elements that we have studied here could, themselves be used to control the layers and geodesic normals if they are put far enough apart; the bump in Fig. \ref{fig:x2plusy2} acts as a converging lens, while the bump in Fig. \ref{fig:xy} acts as a diverging lens. 
The effect of such a compound ``lens'' can be seen in Fig. \ref{fig:compound}. The converging bump on the left leads to the formation of an $x$-axis, downstream grain boundary, as it does in Fig. \ref{fig:x2plusy2}; this particular the grain boundary ends due to the diverging effect in the negatively-curved center region of the bump from Fig. \ref{fig:xy}, but two new diverging grain boundaries appear downstream of this element.
We look forward to a future where ``smectic optometrists'' manipulate the layered order with Gaussian curvature to make novel devices and materials.

\section{A Local Ordering Field from the Extrinsic Geometry}

\subsection{Ground States: Involutes and Evolutes}

In currently available experimental data \cite{hexemerthesis,santangelo07}, the stripes form all at once across the surface as the block copolymers cure.
Similarly, if a nematic on a substrate were cooled into the smectic phase, the formation of striped order might not proceed from left to right as imagined above unless a temperature gradient or a strongly nucleating boundary condition at $x \rightarrow -\infty$ were imposed.
Moreover, in the prior subsections, we have only considered the effect of the intrinsic geometry, embodied in the Gaussian curvature, on the formation and orientation of the layered structure.  This approximation amounts to taking the limit $\lambda_n$ and $\lambda_g$ to zero in Eq. (\ref{eq:energy}). We know from our discussion in the introduction that, when the bending couplings $\lambda_n$ and $\lambda_g$ in Eq. (\ref{eq:energy}) are important, the layers prefer to be perpendicular to the lines of $K=0$ in Figs. \ref{fig:layers}-\ref{fig:xy}.  
Can a smectic texture with equally-spaced layers be found so that it agrees with the preferred direction along lines given by the $K=0$ locus?  In this section we will consider the additional effect of the normal curvature coupling $\lambda_n$ but set $\lambda_g=0$.  The intrinsic curvature term differs essentially from the other two in (\ref{eq:schem}); equal spacing is a local constraint and the extrinsic curvature generates a local preferred direction as illustrated in Eq. (\ref{eq:nelson2}).  However, the intrinsic or geodesic curvature can always be set to zero {\sl locally} because there is a geodesic pointing in every direction at each point of the surface.  
As illustrated in Fig. \ref{fig:gaussbonnet} it is only after evolution normal to the stripes that the
geodesic curvature builds up from zero.  Thus, the geodesic curvature term does not set a {\sl local} directional contstraint.  For this reason we neglect $\lambda_g$ in the following discussion and concern ourselves with the orienting effect arising from $\lambda_n$.  Presumably, local ordering fields would dominate the kinetics, particularly in layer-by-layer growth.

For simplicity we return to the simple Gaussian bump in Fig. \ref{fig:layers}.  First, consider the region for which $K<0$.  Recall from the discussion around Eq. (\ref{eq:kappa}) that there always exists a critical angle $\beta$ relative to the principal directions such that $\kappa_n$ in Eq. (\ref{eq:energy}) vanishes.

As discussed in Secs. I and II, the layers will point in the radial direction along the ring where $K=0$ to minimize the normal bending energy.  Remarkably, it is possible to construct an equally-spaced array of lines satisfying this boundary condition with free boundary conditions at infinity.  This texture is given by the involutes of the ring of zero Gaussian curvature, generalized to a curved surface.  Recall that in flat space, the involutes of a closed plane figure can be generated by wrapping an inextensible string many times around the figure, attaching a pen to the end, and then unwrapping the string under tension \cite{diffgeom}.
In optics, the evolutes are the wave fronts which form a caustic 
-- the patterns which we have shown in Figs.  \ref{fig:layers}-\ref{fig:xy} are the evolutes, generalized to a curved surface, of the straight line at $x=-\infty$.  The generalization is straightforward: we now require that the string lie within the surface so that distances are measured intrinsically.  

\begin{figure}
\includegraphics[width=0.45 \textwidth]{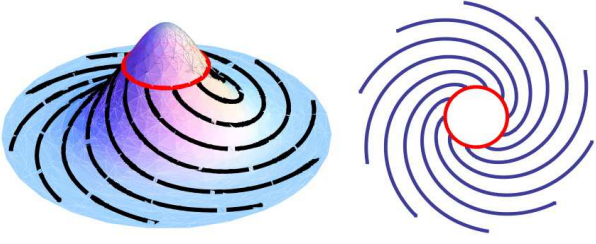}
\caption{This texture of involutes generalized to a curved surface has both uniform layer spacing and minimizes the normal curvature along the circle $K=0$.  Note that the layers bend tightly to be perpendicular to this circle, leading to a diverging curvature.  
 To the right is a top view: a projection onto the $(x,y)$ plane.}
\label{fig:spiral}
\end{figure}

We can make this observation more precise by finding a family of geodesics normal to the layers, from which we can generate uniformly-spaced stripes.  On surfaces of revolution, these geodesics are characterized by Clairaut's theorem \cite{diffgeom},
 $r \sin \theta = r_0$ for a constant $r_0$, where $\theta$ is the angle the geodesic makes with respect to the radial direction. The layers, which lie perpendicular to the normals, therefore make an angle $\psi=\pi/2-\theta$ with the radial direction so $\cos \psi = r_0/r$. As for the bumps discussed above, $r$ is the distance from the axis of revolution; that is, the radial projected distance. When $r = r_0$, corresponding to the circle $K=0$ of the Gaussian bump, $\psi = 0$ and the layers are radial; when $r \rightarrow \infty$, $\psi = \pi/2$ and the layers are azimuthal. Thus, in the absence of defects, the extrinsic alignment forced at $K=0$ leads to a circular ``lines of latitude structure'' for the layers as $r\rightarrow\infty$.  In Fig. \ref{fig:spiral} we show these involutes drawn on the Gaussian bump of Fig. \ref{fig:layers}.  We also note that just as the layer curvature diverges on the caustic cusps in Figs. \ref{fig:layers}-\ref{fig:compound}, the curvature of the layers diverges on the ring.  To see this directly, calculate 
\begin{equation}
\kappa_g = - \nabla \cdot \textbf{n} = - \frac{1}{\sqrt{g}} \partial_i \left[\sqrt{g} n^i\right],
\end{equation}
where the layer normal has components  $n^r = \sin \psi/\sqrt{1+\left(\partial_r h\right)^2}$ and $n^\phi = \cos \psi/r$. For the layers defined above, this yields,
\begin{equation}
\kappa_g = \frac{-1}{\sqrt{1 + \left(\partial_r h\right)^2} \sqrt{r^2-r_0^2}},
\end{equation}
which diverges as $r \rightarrow r_0^+$. Note that our involute {\sl ansatz} interpolates between the minimizer of normal curvature near $K=0$ and the low energy, azimuthal ``bulls-eye'' pattern of the central cone in Fig. \ref{fig:cones}.  Though it might be natural to assume that the bulls-eye pattern would minimize the overall energy, it is frustrated by the orienting normal
curvature energy at the ring of $K=0$ -- experiment also shows that the layers lie perpendicular to this ring \cite{hexemerthesis,santangelo07}. 

Certainly in the case of the involutes, the diverging curvature at $K=0$ will be softened by breaking the zero compressional strain condition.
From dimensional analysis one can argue that at length scales longer than 
$\lambda_n$ and $\lambda_g$, the strain energy dominates the curvature energy and that the $e=0$ approximation is valid.  However, at these cusp-like singularities, the curvature always dominates the energy no matter the scale of 
$\lambda_n$ and $\lambda_g$. The singularities at these cusps are relaxed by the competition between compression and curvature elasticity.    
The proper analysis of the breakdown of equal spacing requires a more complete theory which also allows the formation of edge dislocations in the smectic order \cite{futwork}.

\subsection{Local Order}

Equation (\ref{eq:kappan}), when inserted into Eq. (\ref{eq:energy}), endows the layers with a preferred directionality by minimizing $\kappa_n^2$, and, hence, the normal curvature energy. In regions of positive curvature, the minimal energy direction lies along the principal direction with the smallest curvature. In regions of negative Gaussian curvature, $\kappa_1/\kappa_2 < 0$, so we can solve equation (\ref{eq:kappan}) for $\kappa_n = 0$, finding a preferred angle $\beta_0$ given by (see Eq. (\ref{eq:kappan}))
\begin{equation}\label{eq:normaldirection}
\tan \beta_0 = \sqrt{- \frac{\kappa_1}{\kappa_2}}.
\end{equation}
It is instructive to again consider the case of a cylinder of radius $R$, in which $\kappa_1 = 0$ along the cylinder axis and $\kappa_2 = 1/R$. The preferred angle is $\beta = 0$ which implies that the layers all lie along the cylinder axis. Why this should be so is immediately apparent from Figure \ref{fig:cylinders}: when the layers lie along the cylinder axis they are straight in three dimensions whereas when they are azimuthal, each layer has curvature $1/R$.

On our Gaussian bump,  $h(r) = h_0 e^{-r^2/(2 R^2)}$, and there is a ring of $K = 0$; near this ring, the surface is cylinder-like and we should expect a preferred radial direction for the layers as discussed above. Outside the ring, $K < 0$, and the preferred direction follows equation (\ref{eq:normaldirection}) in a surface dependent way. There is no general behavior that can be inferred via the expressions for the curvatures in Eq. (\ref{eq:kappa}) when $K<0$ -- the precise form of $h(r)$ dictates the ordering direction.  Even
far from the bump there is no universal direction; any angle is possible depending on $\lim_{r \rightarrow \infty} \kappa_r(r)/\kappa_\theta(r)$. 

To probe whether these low curvature directions are actually adopted by the layers, we compute the normal bending energy cost for deviation from the preferred direction for a small patch of layers from Fig. \ref{fig:spiral} in the region where $K<0$.
A straightforward calculation yields,
\begin{eqnarray}
\delta \mathcal{E}  &\approx &\frac{B \lambda_n^2}{2} \left(\kappa_\theta - \kappa_r \right)^2 \int dA~\sin^2 \left(2 \beta_0 \right) \left(\delta \beta\right)^2\nonumber\\&&\qquad=\frac{-B\lambda_n^2K}{2}\int dA~(\delta\beta)^2\label{eq:deltaE}
\end{eqnarray}
when $\beta_0 > 0$.  We have used the condition $\kappa_n=0$ and inserted the preferred angle 
$\beta_0$ in order to simplify Eq. (\ref{eq:deltaE}); the second equality follows from trigonometric identities.  The magnitude $-K>0$ decreases as $r \rightarrow \infty$ but is order $R^{-2}$ up to $r/R \approx 3$. At small angles, however, $\sin (2 \beta_0)$ is small and the ordering field is particularly weak. Near regions where $\beta_0 = 0$ it follows that $\kappa_r=0$ and
\begin{equation}
\delta \mathcal{E} \approx \frac{B \lambda_n^2}{2} \kappa_\theta^2 \int dA~ \left(\delta \beta\right)^4. 
\end{equation}
Thus there is still a weak, anharmonic ordering field even as the layers approach the ring of zero Gaussian curvature, just as for the cylinder described by Eq. (\ref{eq:nelson2}).

\section{Conclusion and Outlook}\label{sec:outlook}

We have outlined a theory of uniformly-spaced lines on a curved surface and used this to understand smectic liquid crystal textures on curved substrates. We started with the more general free energy of Eq. (\ref{eq:energy}), but specialized to the case where the compression energy dominated ($\lambda_g=\lambda_n=0$). We then studied the effect of the weak ordering field embodied in a small $\lambda_n \gg \lambda_g$. This theory is geometrical in nature, and demonstrates that substrate Gaussian curvature subtly frustrates smectic order by bending layers and leading to defects and grain boundaries. We have neglected kinetic effects \cite{GV} focussing on ground state configurations.  

We now comment on additional limitations of our approach for real experimental systems. In particular, two-dimensional smectic order is destroyed by thermal fluctuations and the nucleation of defects \cite{ntsma}. The remaining intermediate-range order leaves a labyrinthine phase of stripes with anisotropic correlation lengths $\xi_{||}$ and $\xi_{\perp}$ that can be considerably larger than a layer spacing, as occurs on a flat substrate. Thus, intrinsic curvature effects (the only effect possible in a flat system) can be weak, and are likely to be less important than extrinsic curvature in determining layer structure. When the effects of normal curvature are compatible with uniform spacing, we expect the layers to be very well-ordered over long length scales precisely because the curvature acts as an ``ordering'' field. This induced intermediate-range order on a corrugated surface does indeed seem evident in experiments of block copolymer cylinders on a bump \cite{santangelo07,hexemerthesis}. In particular, the layers will transition from being radial near the radius of $K=0$ to azimuthal far from the bump. Where bias introduced by normal curvature disagrees with the constraints required by uniform layer spacing, we expect there will be defects in the layers that accommodate the extrinsic curvature effects more closely. Therefore, a more quantitative understanding of the role of defects is necessary to build a complete picture of the ground state of smectic liquid crystals on curved surfaces \cite{futwork}. Our results can serve as a minimal template for smectic textures from which a more detailed theory of defects can emerge.\\

\acknowledgments
We acknowledge stimulating discussions with A. Hexemer, G. Grason, B. Jain, and E. Kramer.  RDK acknowledges the Aspen Center for Physics, where some of this work was completed.  RDK and VV were supported by the National Science Foundation through grant DMR05-47230, via the Penn MRSEC Grant DMR-0520020 and a gift from L. J. Bernstein. Work by DRN was supported by the National Science Foundation, through grant DMR-0654191 and via the Harvard Materials Research Science and Engineering Center through grant DMR-0820484. CDS was supported by the National Science Foundation through DMR-0846852 and the UMass Energy Frontier Research Center through the Department of Energy.

\end{document}